\documentclass[twocolumn,showpacs,prb,superscriptaddress,aps,floatfix]{revtex4-1}
\usepackage{rotating}
\usepackage{amsmath}
\usepackage{amsfonts}
\usepackage{amssymb}
\usepackage{float}

\usepackage[normalem]{ulem}
\usepackage{comment}
\usepackage{color}
\usepackage{graphicx}
\usepackage[active]{srcltx}
\usepackage[sort&compress]{natbib}

\newcommand{\PP}{{\boldsymbol {\cal P}}}
\newcommand{\JJ}{{\boldsymbol {\cal J}}}

\newcommand{\bb}{{\bf b}}

\newcommand{\ra}{\rightarrow}

\newcommand{\zero}{{\bf 0}}
\newcommand{\qq}{{\bf q}}

\newcommand{\rr}{{\bf r}}
\newcommand{\pp}{{\bf p}}
\newcommand{\kk}{{\bf k}}

\newcommand{\jj}{{\bf j}}

\newcommand{\HH}{{\bf H}}

\newcommand{\GG}{{\bf G}}

\newcommand{\w}{\omega}

\newcommand{\chirr}{\chi^{\rho\rho}}
\newcommand{\bchirr}{\bar\chi^{\rho\rho}}
\newcommand{\tchirr}{\tilde\chi^{\rho\rho}}
\newcommand{\be}{\begin{equation}}
\newcommand{\ee}{\end{equation}}
\newcommand{\ben}{\begin{equation*}}
\newcommand{\een}{\end{equation*}}
\newcommand{\bea}{\begin{eqnarray}}
\newcommand{\eea}{\end{eqnarray}}
\newcommand{\bean}{\begin{eqnarray*}}
\newcommand{\eean}{\end{eqnarray*}}

\newcommand{\newtensor}[1] {\underline{\underline{#1}}}

\def\sss{\scriptscriptstyle\rm}
\def\Efield{\boldsymbol{\cal E}}
\def\efield{\boldsymbol{\scriptstyle \cal E}}

\def\susc#1{\underline{\underline{\chi}}^{(#1)}}
\def\PPo#1{{\boldsymbol {\cal P}}^{(#1)}}
\def\tsusc#1{\underline{\underline{\tilde{\chi}}}^{(#1)}}

\newcommand{\cnrs}{CNRS/Univ. Grenoble Alpes, Institut N\'eel, F-38042 Grenoble, France}
\newcommand{\qub}{School of Mathematics and Physics, Queen's University Belfast, Belfast BT7 1NN, Northern Ireland, UK}
\newcommand{\monterotondo}{Istituto di Struttura della Materia of the CNR, Via Salaria Km 29.3, I-00016 Montelibretti, Italy}
\newcommand{\etsf}{European Theoretical Spectroscopy Facilities (ETSF)}
\newcommand{\cinam}{CNRS/Aix-Marseille Universit\'e, Centre Interdisciplinaire de Nanoscience de Marseille UMR 7325 Campus de Luminy, 13288 Marseille cedex 9, France}

\begin{document}
\title{Dielectrics in a time-dependent electric field: a real-time approach based on density-polarization functional theory}

\author{M. Gr\"{u}ning} 
\affiliation{\qub}
\affiliation{\etsf}

\author{D. Sangalli}
\affiliation{\monterotondo}
\affiliation{\etsf}

\author{C. Attaccalite}
\affiliation{\cnrs}
\affiliation{\cinam}
\affiliation{\etsf}

\begin{abstract}
 In the presence of a (time-dependent) macroscopic electric field the electron dynamics of dielectrics cannot be described by the time-dependent density only. We present a real-time formalism that has the density and the macroscopic polarization $\PP$ as key quantities. We show that a simple local function of $\PP$ already captures long-range correlation in linear and nonlinear optical response functions. Specifically, after detailing the numerical implementation, we examine the optical absorption, the second- and third-harmonic generation of bulk Si, GaAs, AlAs and  CdTe at different level of approximation. We highlight links with ultranonlocal exchange--correlation functional approximations proposed within linear response time-dependent density functional theory framework.
\end{abstract}           


\maketitle

\section{Introduction}
Time-dependent density-functional theory\cite{PhysRevLett.52.997} (TD-DFT) is a standard tool in the computation of the optical response of molecules and in general of finite systems. In contrast TD-DFT is rarely employed for the study of the optical response of extended systems such as periodic crystals. The main reason is that within the common approximations TD-DFT fails to describe excitonic effects which typically dominate the optical spectra of insulators and semiconductors.~\cite{botti2007time} 

Though commonly attributed to the approximation for the exchange--correlation (xc) density functional, the problem of TD-DFT for periodic crystals is more fundamental. Calculations of optical response of periodic crystal use periodic boundary conditions. TD-DFT is based on the Runge-Gross theorem~\cite{PhysRevLett.52.997} that establishes the one-to-one correspondence between the time-dependent densities and scalar external potentials. However, for periodic systems in a time-dependent homogeneous electric field only the one-to-one correspondence between the time-dependent currents and potentials (scalar and vector) can be established and time-dependent current density functional theory (TD-CDFT) is then the correct theoretical framework.~\cite{maitra2003current,PhysRevA.38.1149} In particular it is the optical limit, i.e. the case in which the transferred momentum $\qq\ra 0$,  which cannot be described starting from the density only. 
One could still work with functionals that depends on the density-only, but there is a price to pay. All the equations have to be worked out with a finite but very small momentum  and the $\qq\ra 0$ limit can be performed only at the end of the calculation. 
Furthermore in order to describe excitonic effect the xc functionals have to be ultranonlocal and to diverge as $\qq \ra 0$.~\cite{PhysRevLett.88.066404} Such an approach is used within the linear response framework but it is not feasible within a real-time framework since for practical reasons calculations have to be performed directly at $\qq=0$. Thus one needs to go beyond the density-only treatment. As a clear indication of this, the macroscopic polarization and the response functions cannot be calculated within a density-only scheme at $\qq=0$.~\cite{PhysRevB.9.1998}
Problems are not limited to the time--dependent case. Even in the static limit, e.g. for dielectrics in a static homogeneous electric field, Gonze and coworkers proved that {\em ``the potential is not a unique functional of the density, but depends also on the macroscopic polarization''}.~\cite{Gonze1995} In this case then the theory has to be generalized to consider functionals of both the density and the polarization in what is called density--polarization functional theory (DPFT). The latter can be obtained from TD-CDFT in the static limit.

Here we propose a real-time approach based on DPFT for calculating the optical response properties of dielectrics, thus considering functionals of both the time-dependent density and the macroscopic bulk polarization.
Real-time approaches allows in principle to calculate the optical response at all order so to access nonlinear properties,\cite{takimoto:154114} including nonperturbative extreme nonlinear phenomena\cite{lee2014first} and to simulate real-time spectroscopy experiments.\cite{otobe2015femtosecond} It is highly desirable then to have computational inexpensive first principles real-time approaches, such as TD-DFT, that include excitonic effects.            
In particular here we consider an effective electric field which is a functional of the macroscopic polarization. We employ simple local functionals of the polarization~\cite{maitra2003current,PhysRevLett.115.137402,DeBoeij2001} either fitted to reproduce the linear optical spectra\cite{botti2004long} or derived from the jellium with gap model kernel.~\cite{jgm}

In the following, we review DPFT and we extend it to the case of time-dependent electric fields. We discuss briefly the approximations for the effective electric field and we present how the relevant response functions are calculated from the macroscopic polarization.  Then, we show that for the optical absorption, the second-harmonic generation (SHG) and third-harmonic generation (THG) of semiconductors  the simple local functionals of the polarization account for excitonic effects similarly to the ultranonlocal kernel within the density-only response framework. In the conclusion we discuss the proposed approach as an alternative to existing schemes based on TD-DFT and TD-CDFT.

\section{Density polarization functional theory}

The coupling of an external electromagnetic field with a dielectric is described via the electromagnetic potentials and thus is gauge dependent.
In the present manuscript we use the length gauge which is obtained from the multipolar gauge within the electric-dipole approximation (EDA).~\cite{Kobe1982}. 
This implies that we assume a spatially uniform electric field. Such macroscopic electric field enters via a scalar potential,
$\varphi(\rr)=-\Efield^{\text{ext}} \cdot \rr$, whose corresponding energy has generally the form
\be
E_\varphi= -\Omega\, \Efield^{\text{ext}} \cdot \PP 
\label{eq:enphi}
\ee
where $\Omega$ is the volume and $\PP$ is the bulk macroscopic polarization that is then the key quantity to describe the coupling of dielectrics with external fields 
(in the velocity gauge the coupling would have been instead via the macroscopic current). 

For finite systems (i.e. in which the electronic density $n$ goes to zero when $\rr\rightarrow \infty$), Eq.~\ref{eq:enphi} is equivalent to 
$\int n(\rr) \varphi(\rr) d\rr$ and $\PP = \int n(\rr) \rr d\rr$.  
However these expressions are ill-defined when periodic boundary conditions are imposed.~\cite{Blount} The Modern Theory of Polarization~\cite{RevModPhys.66.899} provides a correct definition for the macroscopic bulk polarization in terms of the many-body geometric phase.
For a system of independent particles in a periodic potential the polarization ${\cal \PP}$ along the Cartesian direction $\alpha$ is given by~\cite{KSV1} 
\be
{\cal P}_\alpha = -\frac{ief}{(2 \pi)^3} \sum_{n}^{\text{occ}} \int d\kk \langle  u_{\kk n} |\partial_{k_\alpha}  u_{\kk n} \rangle, \label{berryP} 
\ee
where $f$ the spin occupation, and $|u_{\kk n}\rangle$ the periodic part of the Bloch function $|\phi_{\kk n}\rangle$.  

Equation~(\ref{berryP}) seems to suggest that, though $\PP$ cannot be expressed as an explicit functional of the electron density $n$, it is still an implicit functional of $n$ through the Bloch functions obtained from the solution of the Kohn-Sham (KS) equation. As we discuss in the following subsection however, for a dielectric in a macroscopic electric field the macroscopic polarization needs to be considered as an independent variable. Accordingly
the macroscopic part of the external electric field $\Efield^{\text{ext}}$ cannot be included via the potential $v^{\text{ext}}$,
since the associated energy functional would be ill-defined. In such approach the KS equations and the associated Bloch functions depend on both the density and the macroscopic
polarization of the system.

In the following we use the Gaussian system of units (or cgs) for the polarization, electric fields and the susceptibilities. 

\subsection{Static case}
An infinite periodic crystal (IPC) in a macroscopic electric field $\Efield^{\text{ext}}$ does not have a ground-state. Therefore the Hohenberg-Kohn theorem cannot be applied and DFT cannot be used. In particular the density does not suffice to describe the system as the one-to-one mapping between density and external potential does not hold: one can devise an external macroscopic electric field that applied to a system of electrons in an IPC does not change its density $n$.
The works of Gonze Ghosez and Godby,~\cite{Gonze1995} Resta~\cite{Resta1996}, Vanderbilt~\cite{Vanderbilt1997} and of Martin and Ortiz~\cite{Martin1997} established that in addition to the density, the macroscopic (bulk) polarization $\PP$ is needed to characterize IPC in a macroscopic electric field. With some cautions the proof of the Hohenberg-Kohn theorem can be extended~\cite{Martin1997} to demonstrate the existence of the invertible mapping
$$(n(\rr),\PP)\leftrightarrow (\bar v^{\text{ext}}(\rr),\Efield^{\text{ext}}) $$
where $ \bar v_{\text{ext}}$ is the periodic microscopic part of the external potential.
Then the total energy of an IPC is a functional of both the electron density $n$ and the macroscopic polarization $\PP$:
\be \label{eq:dpften}
E[n,\PP] = \bar F[n,\PP]+ \int_\Omega n(\rr) \bar v^{\text{ext}}(\rr)\,d\rr  -\Omega\, \Efield^{\text{ext}} \cdot \PP,
\ee 
where $\bar F$, the internal energy, is a universal functional of both $n$ and $\PP$ (see Ref.~\onlinecite{Martin1997} for details).
and is defined in the usual way
as the sum of the expectation the kinetic and electron-electron interaction operators
\be
\bar F[n,\PP] = \langle \Psi | \hat T + \hat V_{ee} |\Psi \rangle.
\ee 
The difference with the internal energy within standard DFT is that the $N$-particle wavefunction $\Psi$ is not an eigenstate of the original Hamiltonian (which does not have a ground state), but of an auxiliary Hamiltonian which commutes with the translation operator (see Ref.~\onlinecite{Martin1997} for details). Notice that DPFT is not the only way to treat IPC in a electric field within a density functional framework: as an alternative Umari and Pasquarello proposed $\Efield$-DFT, a density functional theory depending on the electric field.~\cite{Umari2005}   

The KS equations can be extended as well to treat IPC in a macroscopic electric field.~\cite{Martin1997}
In particular the KS crystal Hamiltonian takes the form:
\be \label{eq:dpftks}
H^{s}_\kk
= -\frac{1}{2}\left(\nabla + i\kk\right )^2 + \bar v^{s}(\rr) -\Omega\Efield^{s}\cdot \nabla_\kk 
\ee
which is a functional of both the density and the polarization.
In Eq.~\eqref{eq:dpftks}  the KS microscopic (periodic) potential $\bar v^{s}$ is defined as
\be\label{eq:kspot}
\bar v^{s}(\rr) = \bar v^{\text{ext}}(\rr) +  \bar v^{\text{H}}(\rr) + \bar v^{\text{xc}}(\rr) 
\ee
$\bar v^{\text{ext}}(\rr)$, $\bar v^{\text{H}}$ are respectively the microscopic external and Hartree potential.
The total classical potential is defined as $\bar v^{\text{tot}}(\rr)=\bar v^{\text{ext}}(\rr)+\bar v^{\text{H}}(\rr)$.
$\bar v^{xc}$ is the functional derivative of the xc energy with respect to the density.
$\bar v^{\text{ext}}(\rr)$ here describes the field generated by the ions, i.e. the electron--ion interaction
in the Coulomb gauge and neglecting retardation effects.
The last term of the RHS of Eq.~\eqref{eq:dpftks}---that originates from the last term in the RHS of Eq.~\eqref{eq:dpften}---constitutes
the key difference with respect to the zero-field KS equations.
$\nabla_\kk$~is the polarization operator derived by functional-differentiating
$\PP$ [Eq.~\eqref{berryP}] with respect to the KS eigenstates.
$\Efield^{s}$ is the KS macroscopic field  
\be
\Efield^{s} = \Efield^{\text{ext}} + \Efield^{\text{H}}  + \Efield^{\text{xc}},
\label{eq:ksfld}
\ee
that contains the corresponding macroscopic components of $\bar v_s$. Note that these macroscopic components cannot be included via the potential which would be ill-defined when imposing periodic boundary conditions.  
The $\Efield^{\text{xc}}$ is defined as the partial derivative of the xc energy with respect to the polarization density field.
The sum of the macroscopic external and Hartree fields defines the total classical field:
\be
\Efield^{\text{tot}} = \Efield^{\text{ext}} + \Efield^{\text{H}}. 
\label{eq:totfld}
\ee    

At zero-field, that is when no macroscopic external electric field $\Efield^{\text{ext}}$ is applied,
the macroscopic component of the ionic potential and of the Hartree component exactly cancel as a consequence of the charge neutrality of the system and  
the macroscopic xc component vanishes.
In this situation  standard density-only functional theory can be used.

As $\bar v^{s}$ and $\Efield^{s}$ are functionals of the density and the polarization, the KS equations
for the KS orbitals $\{\phi_{n\kk}\}$  have to be solved self-consistently with the density (spin unpolarized case)
\be
n(\rr) = 2 \sum^{\text{occ}} |\phi_{n\kk}(\rr)|^2 
\ee
and the polarization expressed in terms of a Berry phase [Eq.~\eqref{berryP}].

\subsection{Time-dependent case}
The Runge-Gross theorem \cite{PhysRevLett.52.997} is the basis of TD-DFT. It establishes the one-to-one mapping between the time-dependent scalar potential and the time-dependent density. For the case in which a time-dependent vector potential is present Ghosh and Dhara~\cite{PhysRevA.38.1149} showed that the mapping can be established between the current-density and the vector potential. More recently Maitra and co-workers~\cite{maitra2003current} showed that TD-CDFT is the correct framework for IPC in homogeneous electric fields.

The time-dependent change in the polarization density field $\pp$ is related to the time-dependent current-density $\jj$ by 
\be
\pp (\rr;t) = \int^t_{-\infty} dt' \jj(\rr;t') 
\label{eq:curpol}
\ee
In a dielectric we can then use either $\pp(\rr,t)$ or $\jj(\rr, t)$ as main variable to describe an IPC in a time-dependent finite homogeneous electric field. Furthermore we can consider separately the microscopic and the macroscopic components of $\pp(\rr, t)$: $\PP (t)$ and $\bar \pp(\rr, t)$. The longitudinal component of latter quantity is determined by the density through the continuity equation. When interested in the optical limit (and working in the EDA), the microscopic transverse component can be neglected\cite{noteRod} and we can extend to the time-dependent case the one-to-one mapping.
$$(n(\rr,t),\PP(t))\leftrightarrow (\bar v^{\text{ext}}(\rr,t),\Efield^{\text{ext}}(t)). $$
The time-dependent Kohn-Sham crystal Hamiltonian
has the same form of the equilibrium KS Hamiltonian:
\be \label{eq:tdksh}
H^{s}_\kk(t)
= -\frac{1}{2}\left(\nabla + i\kk\right )^2 +  \bar v^{s}(\rr,t) -\Omega\, \Efield^{s}(t)\cdot \nabla_\kk
.
\ee

We rewrite the external field and potential\cite{notevrep} as the contribution at equilibrium, $\Efield^{\text{ext},0}$ and $\bar v^{\text{ext},0}(\rr)$  plus the time-dependent perturbation:
\bea
\Efield^{\text{ext}}(t)&=&\Efield^{\text{ext},0}+\Delta \Efield^{\text{ext}}(t), \\
\bar v^{\text{ext}}(\rr,t)&=&\bar v^{\text{ext},0}(\rr) +\Delta v^{\text{ext}}(\rr,t) .
\eea  
Then,
\bea
\bar v^{s}(\rr,t) &=& v^{s,0}(\rr) +\Delta  \bar v^{s}(\rr,t) \\
\Efield^{s}(t)     &=&\Efield^{s,0}+\Delta \Efield^{s}(t),
\eea
where the $0$ superscript denotes that the functional is evaluated 
in presence of the equilibrium fields, thus at equilibrium density and polarization.  We then restrict ourselves to consider  the case with no external macroscopic electric field at equilibrium, i.e. ${\Efield^{\text{ext},0}=\mathbf{0}}$, and to a macroscopic-only time dependent perturbation, i.e. $\Delta \bar v^{\text{ext}}(\rr,t)=0$.
Therefore
\bea
\Delta \bar v^{s}(\rr, t ) &=& \Delta\bar v^\text{H}+\Delta\bar v^{\text{xc}} \\
\Delta\Efield^{s}(t)&=&\Efield^{s}(t)
\eea

Finally, the TD-KS equations for the periodic part $u_{n\kk}$ of the Bloch function 
can be expressed as
\be
  i\partial_t u_{n\kk} =\left( H^{s,0}_\kk +  \Delta \bar  v^{s}(\rr,t) -\Omega\, \Efield^{s}(t)\cdot \nabla_\kk \right)u_{n\kk}
,
\label{eq:kseom}
\ee
and have to be solved consistently with the time-dependent density and polarization.
The latter has the same form of the static polarization [Eq.~\eqref{eq:kseom}] with
the difference that $|v_{\kk n}\rangle$ are the time-dependent valence bands.~\cite{souza_prb} 

In the time dependent case and within the EDA, it can be shown straightforwardly that the Hamiltonian in Eq.~\eqref{eq:tdksh} can be derived from the KS Hamiltonian of TD-CDFT with a gauge transformation from the velocity to the length gauge.\cite{maitra2003current}

\section{Expressions for the Kohn-Sham electric field}
\label{sc:ksef}
The KS electric field in Eq.~\eqref{eq:ksfld} is the sum of three components. It seems natural to consider the external component $\Efield^{\text{ext}}$ as an input of the calculation, i.e. $\Efield^{\text{ext}} = \Efield^{\text{inp}}$. The total classical field $\Efield^{\text {tot}}$ is then calculated from Eq.~\eqref{eq:totfld} by adding the Hartree component that in the EDA is the polarization $\Efield^{H} = 4\pi \PP$. This is not the only possible choice nor always the most convenient. When calculating linear and nonlinear optical susceptibilities, which do not depend on the total or external fields, it is numerically more convenient to choose the total classical field as input field. As this work objective is the calculations of optical susceptibilities we adopt indeed $\Efield^{\text{inp}} = \Efield^{\text{tot}}$.  The two choices for the input field, i.e. either the total or external field, have been referred as ``longitudinal geometry'' and ``transverse geometry'' by Yabana and coworkers\cite{PhysRevB.85.045134} and are discussed in more length in Appendix~\ref{appA}. 

While the choice of the input field is a matter of computational convenience, the choice of the expression for the xc macroscopic electric field is critical to the accuracy of the results. Like the microscopic xc potential no exact expression is known and one should resort to an approximation for the functional form of the xc field. Contrary to the microscopic xc potential for which hundreds of approximations exist,\cite{libxc} except for the work of Aulbur and coworkers~\cite{aulbur1996polarization} we are not aware of approximations for the xc macroscopic field. What does exist in the literature are xc kernels within the TD-DFT and TD-CDFT that give a non-zero contribution to the response in the optical limit. In what follows we link the xc kernel with the macroscopic field (similarly to Refs.~\onlinecite{maitra2003current,PhysRevLett.115.137402}). 
In fact in the linear response limit the xc electric field is related to the polarization $\pp$ (see for example Refs.~\onlinecite{maitra2003current,PhysRevLett.115.137402,DeBoeij2001})
through the xc kernel $\newtensor{F}^{\text{xc}}_{}$. The latter describes how the xc electric field (both microscopic and macroscopic) changes when the polarization is perturbed. $\newtensor{F}^{\text{xc}}_{}$ can be defined independently through the Dyson equation connecting 
the polarization response function of the physical system, $\newtensor{\chi}$,
to the polarization response function of the KS system, $\newtensor{\chi}^{s}$. 
By rewriting the relation between $\Efield^{\text{xc}}$ and $\newtensor{F}^{\text{xc}}_{}$ in reciprocal space~\cite{notepath} one obtains~\cite{maitra2003current} for the macroscopic component ($\GG=0$) 
\begin{multline}
\Efield^{\text{xc}}(t)= \int dt' \Big[ \newtensor{F}^{\text{xc}}_{\sss 00}(t-t') \PP(t')  \\
               -{i} \sum_{\GG' \neq 0} \newtensor{F}^{\text{xc}}_{\sss 0\GG'}(t-t')\frac{ n_{\sss \GG'}(t')}{{\rm G'}^2}\GG' \Big] 
\label{Excmac}
\text{,}
\end{multline}
and for the microscopic components $\efield^{\text{xc}}_{\sss \GG}$ ($\GG\neq 0$)
\begin{multline}
\efield^{\text{xc}}_{\sss \GG}(t)= \int dt' \Big[ \newtensor{F}^{\text{xc}}_{\sss \GG 0}(t-t') \PP(t')  \\
                    -{i} \sum_{\GG' \neq 0} \newtensor{F}^{\text{xc}}_{\sss \GG\GG'}(t-t')\frac{ n_{\sss \GG'}(t')}{{\rm G'}^2}\GG' \Big] 
\label{Excmic}
\text{.}
\end{multline}
The first term on the RHS of Eq.~\eqref{Excmac} is directly proportional to the macroscopic polarization, the second term involves the density and is the microscopic contribution to the macroscopic field. Note that as we assume the EDA we do not have the contribution from the microscopic transverse current as in Maitra and coworkers.~\cite{maitra2003current}
The variation of the microscopic xc potential $\Delta \bar v^{\text{xc}}$ can be written in terms of the microscopic components $\efield^{\text{xc}}_{\sss \GG}$ as 
\be
\Delta \bar v^{\text{xc}}_{\sss \GG}(t) =  i \frac{\GG \cdot  \efield^{\text{xc}}_{\sss \GG}(t) }{{\rm G}^2}.
\label{eq:pot_from_exc}
\ee
  
Berger~\cite{PhysRevLett.115.137402} has recently proposed an approximation for $\newtensor{F}^{\text{xc}}_{00}$ from TD-CDFT. The approximation however requires the knowledge of the Random-Phase-approximation (RPA) static dielectric function: while within a linear response approach this does not require any additional calculation, within a real-time approach the RPA static dielectric function needs to be computed beforehand. Previously, again within TD-CDFT, de Boeij and coworkers~\cite{DeBoeij2001} had derived an approximation for the  $\newtensor{F}^{\text{xc}}_{00}$ from  the Vignale-Kohn current-density functional\cite{PhysRevLett.77.2037}. Both these approximations successfully describe long range effects in optical absorption spectra of dielectrics.   

An alternative way to derive approximations for $\newtensor{F}^{\text{xc}}_{}$ is to rely on the standard TD-DFT xc kernel $f^{\text{xc}}_{}$.\cite{noterev} The latter describes how the xc potential changes when the density is perturbed and is defined from the Dyson equation relating the density-density response of the physical and the KS system. The general relation between $\newtensor{F}^{\text{xc}}_{}$ and $f^{\text{xc}}_{}$ ``involves repeated inversions of tensor integral operators''\cite{PhysRevB.76.205103} and it is not of practical use. In the long wavelength limit this relation simplifies and the two kernels can be related via the equation,\cite{maitra2003current}
\be 
f^{\text{xc}}_{\sss \GG \GG'}(\qq\rightarrow 0; t - t') = \lim_{\qq\rightarrow 0} \frac{\newtensor{F}^{\text{xc}}_{\sss \GG \GG'}(t-t') \cdot \newtensor{g}}{|\qq + \GG||\qq + \GG'|}.  
\label{eq:k2k}
\ee 
where $\newtensor{g}$ is the metric tensor. For example the long-range corrected (LRC) approximations $f^{\text{xc}}\approx f^{\sss LRC}$, which take the form
\be
f^{\sss LRC}_{\sss \GG\GG'}(\qq\rightarrow 0; t-t') =
  \lim_{\qq\rightarrow 0} \frac{-\alpha^{\sss LRC}}{|\qq|^2}\delta_{\sss \GG,\zero}\delta_{\sss \GG',\zero}\delta(t-t'),   
\label{jgmqzero}
\ee
can be used (we assume here and it what follows $\alpha > 0$). Then $\newtensor{F}^{\text{xc}}_{00} \cdot \newtensor{g} \approx -\alpha^{\sss LRC}$. 

In this work, we derive the $\newtensor{F}^{\text{xc}}$ needed in Eq.~\eqref{Excmac} from the Jellium with Gap Model (JGM) kernel proposed by Trevisanutto and coworkers.\cite{jgm}
The latter kernel is a functional of the electronic density $n$ and of the fundamental gap of the material $E_{\text{gap}}$. In the optical limit the JGM kernel takes the form of a long-range corrected approximation
with $\alpha^{\sss LRC}$ defined as the cell average~\cite{jgm} of
\be
\alpha^{\sss JGM}(\rr;t) = 4\pi \tilde B \left[1 - \exp{\left( -\frac{E_{\text{gap}}^2}{4\pi n \tilde B} \right)} \right].  
\label{eq:alpha}
\ee
In the equation above $\tilde B = (B + E_{\text{gap}})/(1 + E_{\text{gap}})$, where $B=B[n]$ is a functional of the density found by fitting the local field factor of the homogeneous electron gas from Quantum Montecarlo data.~\cite{PhysRevB.57.14569}   
For cubic systems we thus consider $\newtensor{F}^{\text{xc}}\approx\newtensor{F}^{\sss JGM}$ with
\begin{subequations}
\begin{gather}
\newtensor{F}^{\sss JGM}_{\sss \zero\GG}(t-t') = -\frac{1}{2}\alpha^{\sss JGM}_{\sss \GG}(t) \newtensor I \delta(t-t') \\
\newtensor{F}^{\sss JGM}_{\sss \GG\zero}(t-t') = -\frac{1}{2}\alpha^{*\sss JGM}_{\sss \GG}(t) \newtensor I \delta(t-t').
\end{gather}
\label{eq:krnl}
\end{subequations}
where $\alpha^{\sss JGM}_{\sss \GG}(t)$ is the Fourier transform of Eq.~\eqref{eq:alpha} and we restricted ourselves to cubic systems for which the metric tensor is the identity, $\newtensor I$. This latter restriction is not fundamental and the above equations can be generalized straightforwardly to non-cubic systems.  
Notice that we symmetrized ${F}^{\sss JGM}_{\sss \GG,\GG'}$ so to obtain a Hermitian kernel. Other strategies of symmetrization have been proposed in the literature, see Ref.~\onlinecite{jgm} and reference therein.

Like standard approximations for the xc kernel this approximation neglects memory effects (i.e. the macroscopic field at time $t$ depends on the values of the density and polarization only at time $t$) and it is thus frequency independent. Several frequency dependent approximations have been derived from current-density functional theory~\cite{PhysRevLett.115.137402,DeBoeij2001,PhysRevLett.102.113001,Nazarov2010}. Contrary to approximations for the $\alpha^{\sss LRC}$ proposed in the literature so far, the derived approximation for $\alpha$ depends on the reciprocal lattice vectors. Furthermore at difference with the approximations proposed in Refs~\onlinecite{botti2004long,PhysRevB.72.125203} this approximation does not rely on empirical parameters---similarly to the family of bootstrap kernels~\cite{PhysRevLett.114.146402,PhysRevLett.107.186401} (that relate $\alpha$ to the electronic screening in an expression equivalent to that derived by Berger from TD-CDFT).

Inserting the approximation for the kernel [Eq.~\eqref{eq:krnl}] in the expression for the xc fields [Eq.~\eqref{Excmac}--\eqref{Excmic}] and using Eq.~\eqref{eq:pot_from_exc} we obtain 
\bea
\Efield^{\sss JGM}(t)&=&\alpha^{\sss JGM}_{\zero}(t)
{\cal \PP}(t)
-\frac{i}{2}\sum_{\sss \GG \neq 0} \alpha^{\sss JGM}_{\sss \GG} (t)\frac{ n_\GG(t)}
{{\rm G}^2}\GG \nonumber \\
\Delta \bar v^{\sss JGM}_{\sss \GG}(t)&=&\frac{i}{2}\sum_{\sss \GG \neq 0}  \frac{\alpha^{*\sss JGM}_{\sss \GG} (t)}{G^2} \GG \cdot {\cal \PP}(t), 
\label{eq:Excapp}
\eea
where the second term in the RHS of Eq.~\eqref{Excmic} is zero due to our symmetrization strategy [Eq.~\eqref{eq:krnl}]. 

In our calculations we will use either Eq.~\eqref{eq:Excapp} and or the empirical $\alpha^{\text{opt}} \PP$ approximation for the macroscopic xc electric field in which $\alpha^{\text{opt}}$ is a parameter which gives the best agreement between the computed and experimental optical absorption spectra. The two approximations will be referred as JGM polarization function (JGM-PF) and optimal polarization functional (opt-PF).

\section{Computational details}

The eigensolutions $\{|\phi^0_{m\kk}\rangle \}$ of the zero-field Hamiltonian are calculated using the plane-wave pseudopotential density-functional code {\sc abinit}~\cite{abinit} within the local density approximation for the xc energy. The kinetic cutoff, the lattice constant and the components included in the valence and type of the pseudopotential  used in these calculations are collected in Table~\ref{tb:pardft}. 
We have employed norm-conserving pseudopotentials of the Troullier-Martins type~\cite{PhysRevB.43.1993} for Si, AlAs and CdTe, and of the Hamann type~\cite{PhysRevB.40.2980} for GaAs. For all the systems we have used four shifted $8 \times 8 \times 8$ Monkhorst-Pack meshes\cite{PhysRevB.13.5188} to converge the ground-state density.   
The periodic part $\{|u^0_{m\kk}\rangle\}$  of the so-generated eigensolutions are used as a basis to expand the time-dependent KS Bloch-functions (or more precisely their periodic part)
\be
|u_{n\kk}(t) \rangle = \sum_m |u^0_{m\kk} \rangle \langle u^0_{m\kk}|u_{n\kk} (t) \rangle =  \sum_m  |u^0_{m\kk} \rangle c^{\kk}_{mn}(t)
\label{eq:basis}
\ee
and the TD-KS equations [Eq.~\eqref{eq:kseom}] can be rewritten as the equation of motions for the coefficients $c^{\kk}_{nm}$. We obtained converged spectra by truncating the sum in Eq.~\eqref{eq:basis} at $m=9$ bands for Si, $m=11$ bands for AlAs and GaAs, and $m=13$ for CdTe. 

\begin{table}[H]
\begin{tabular}{l|c|c|l|l}\hline\hline
  System & $K$ (Ha)& $a_{\text{latt}}$ (Bohr) & atom$_1$ & atom$_2$\\
  \hline
  Si  &  14 &  10.260 & Si: $\ 3s^23p^2$  &\\
  GaAs&  30 &  10.677 & Ga: $4s^24p^1$    & As: $4s^24p^3$\\
  AlAs&  20 &  10.696 & Al: $\ 3s^23p^1$  & As: $4s^24p^3$\\
  CdTe&  40 &  12.249 & Cd: $4d^{10}5s^2$ & Te: $4d^{10}5s^25p^4$\\
  \hline\hline
\end{tabular}
\caption{\footnotesize{Parameters for the DFT calculations.
The kinetic energy cutoff $K$,
the lattice constant $a_{\text{latt}}$ and
the non frozen electrons explicitly included in valence.}} \label{tb:pardft}
\end{table}

The derivatives with respect to the crystal momentum that appear in Eqs.~\eqref{berryP}  and~\eqref{eq:dpftks} for the polarization and the polarization operator are evaluated numerically. Following Souza and coworkers~\cite{souza_prb} the polarization is rewritten as
\be
{\cal P}_\alpha = -\frac{ef}{2 \pi \Omega} \frac{\mathbf a_\alpha}{N_{\kk_\alpha^\perp}} \sum_{\kk_\alpha^\perp} \mbox{Im ln} \prod_{i=1}^{N_{\kk_\alpha}-1}\ \mbox{det } S(\kk_i , \kk_i + \Delta \kk_\alpha). \label{numBP}
\ee
where $\Omega$ is the cell volume, $a$ is the lattice vector, $N_{\kk_\alpha^\perp}$ is the number of $\kk$-points in the plane perpendicular to reciprocal lattice vector $\bb_\alpha$
and $\Delta k_\alpha$ the spacing between two successive $\kk$ points in the $\alpha$ direction. $S$ is the overlap matrix
\be
S_{mn}(\kk , \kk + \Delta \kk_\alpha) = \langle u_{m\kk} | u_{n\kk + \Delta\kk_\alpha} \rangle. 
\label{eq:ovrlap}
\ee
The field coupling operator $\hat w_\kk = \Efield\cdot\nabla_\kk$ is calculated as 
\begin{equation}
        \hat{\rm w}_{\kk}(\boldsymbol{\cal E}) = \frac{ie}{4\pi} \sum_{i=\alpha}^3\,N_\alpha^\parallel (\Efield \cdot {\bf a}_\alpha) \frac{ 4 D(\Delta \kk_\alpha) - D( 2 \Delta \kk_\alpha)}{3}, 
\label{eq:wkhat2}
\end{equation}
where $N_{\kk_\alpha^\parallel}$ is the number of $\kk$-points along the reciprocal lattice vector $\bb_\alpha$ and 
\bea
D(\Delta \kk_\alpha) &=& \frac{1}{2} \left (\hat{P}_{\kk_i + \Delta \kk_\alpha} - \hat{P}_{\kk_i - \Delta \kk_\alpha} \right ), \label{eq:wkhat} \\
\hat{P}_{\kk_i + \Delta \kk_\alpha} &=& \sum_n^{\text{occ}} | \tilde u_{n \kk_i + \Delta \kk_\alpha} \rangle\langle u_{n \kk_i}| \label{eq:proj}
\eea
In the definition for the projector [Eq.\eqref{eq:proj}] $| \tilde u_{n \kk_i + \Delta \kk_\alpha} \rangle$ are gauge-covariant,\cite{souza_prb} i.e. are constructed so that transform under unitary transformation in the same way as $|u_{n \kk_i} \rangle$:
\be
| \tilde u_{n \kk_i + \Delta \kk_\alpha} \rangle = \sum_m^{\text{occ}} [S^{-1}(\kk , \kk + \Delta \kk_\alpha)]_{mn} | u_{m \kk_i + \Delta \kk_\alpha} \rangle.
\ee
Equation~\eqref{eq:wkhat2}, proposed by Nunes and Gonze,~\cite{gonze} corresponds to approximate the Gauge covariant derivative in Eq.~\eqref{eq:kseom} with a finite difference five-point midpoint formula. The truncation error in this expression converges as ${\cal O}(\Delta \kk^4)$ whereas the three-point midpoint formula proposed in Ref.~\onlinecite{souza_prb} and used in our previous works~\cite{nloptics2013,PhysRevB.89.081102,attaccalite2015strong} converges as ${\cal O}(\Delta \kk^2)$. Though more cumbersome, we prefer Eq.~\eqref{eq:wkhat2}, since we noticed that when using polarization dependent functionals the equations of motion (EOMs) are very sensitive to numerical error.
To converge the spectra we considered $24\times24\times 24$ mesh for Si and GaAs, $18\times 18 \times 18$ for AlAs and CdTe. 

In the TD-KS equation [Eq.~\eqref{eq:kseom}] we introduce a phenomenological dephasing by adding a decay operator
\be
R_{n\kk}(t) = \frac{1}{\tau_{n\kk}} \left \{ | u_{n\kk}(t)\rangle \langle  u_{n\kk}(t)| - | u_{n\kk}^0\rangle \langle u_{n\kk}^0| \right \} 
\ee
where the dephasing time $\tau$ can depend on the band and crystal momentum indices.
Those parameters take into account memory-effects from missing electron correlation and from the coupling with the ``environment'' (e.g. defects, phonons) that eventually lead to the finite lifetime of the excitation. Those parameters can be in principle obtained from theory, for example in the context of Green's function theory they can be obtained from the imaginary part of a  self-energy.  Here we choose a dephasing time $\tau$ independent from the band and crystal momentum indices in such a way to reproduce the broadening of the experimental spectrum. For the nonlinear optical spectra we used a broadening of 0.2~eV equivalent to a dephasing time of 6.58~fs. For the absorption spectra we used a broadening of 0.02~eV equivalent to a dephasing time of about 60~fs, and in the post-processing we applied a further Gaussian broadening of 0.1~eV.

We introduce as well a scissor operator $\Delta H^{QP}_\kk$ to correct the KS band gap.
The value of the scissor correction can be calculated from first principles
(e.g. from $GW$ calculations~\cite{aryasetiawan1998gw}), but in this work we choose the
correction so to reproduce the band gap values found in the literature (Table~\ref{tb:lrcprm}). Table~\ref{tb:lrcprm} reports further the optimal value for $\alpha$ used in the opt-PF approximation as suggested by Botti and co-workers\cite{botti2004long}.  For CdTe---for which to our knowledge there are no time-dependent DFT calculations with the LRC kernel---we use 0.2 which is obtained from the fit proposed in Ref~\onlinecite{botti2004long} to extract the optimal $\alpha$ from the experimental dielectric constant.~\cite{singh1993physics}

The final EOM is thus
\be
i\partial_t | u_{n\kk} (t)\rangle =
  \left[ \HH^{s}_\kk(t)+\Delta H^{QP}_\kk + i R_\kk(t)\right] 
     | u_{n\kk}  \rangle \label{eq:def1} .
\ee

We perform real-time simulations using a development version of {\sc Yambo}~\cite{yambo}. For the nonlinear optical properties we input a weak monochromatic electric field for a comb of frequencies in the range of interest and we obtain the frequency dependent response functions from the polarization by Fourier inversion formula (see Ref.~\onlinecite{nloptics2013} for details and  App.~\ref{appA}). For the linear optical properties we input a delta like pulse and obtain the frequency dependent response from the polarization by Fourier transform. The EOMs are integrated using the numerical method proposed in Ref.~\onlinecite{souza_prb} and used in previous works~\cite{nloptics2013,PhysRevB.89.081102}  with a time-step 0.01~fs.  

 \begin{table}[H]
 \begin{tabular}{l|cccc}\hline\hline
   Par/Sys &Si& GaAs& AlAs& CdTe\\
   \hline
   $\alpha$      & 0.2&  0.2 & 0.35 & 0.2 \\
   $\Delta$ (eV) & 0.6&  0.8 & 0.9 &  1.0\\
   \hline\hline
 \end{tabular}
\caption{\footnotesize{Material dependent parameters used in the simulations: the parameter $\alpha$ employed in the opt-PF approximation and the value of the scissor operator.}} \label{tb:lrcprm}
 \end{table}

\section{Results} \label{rt-tddft}

We considered the optical properties of bulk Si, which has a diamond structure, and GaAs, AlAs and CdTe, which have zincblende structure. The two structures are similar, both are face-centered cubic systems with a two atom basis (at the origin, and at 1/4 of the unit cell in each direction). In silicon the two atoms are identical, in the zincblende structures are the different atoms of the II-VI (CdTe) or III-V (GaAs and AlAs) compound. In terms of crystal symmetries this implies that at variance with silicon they miss the inversion symmetry, and therefore have a dipole-allowed SHG. In what follows we study linear and nonlinear optical properties contrasting the standard time-dependent local density approximation (TD-LDA) with the real-time DPFT approach.

\subsection{Optical absorption}

\begin{figure}[t]
\centering
\includegraphics[width=0.5\textwidth]{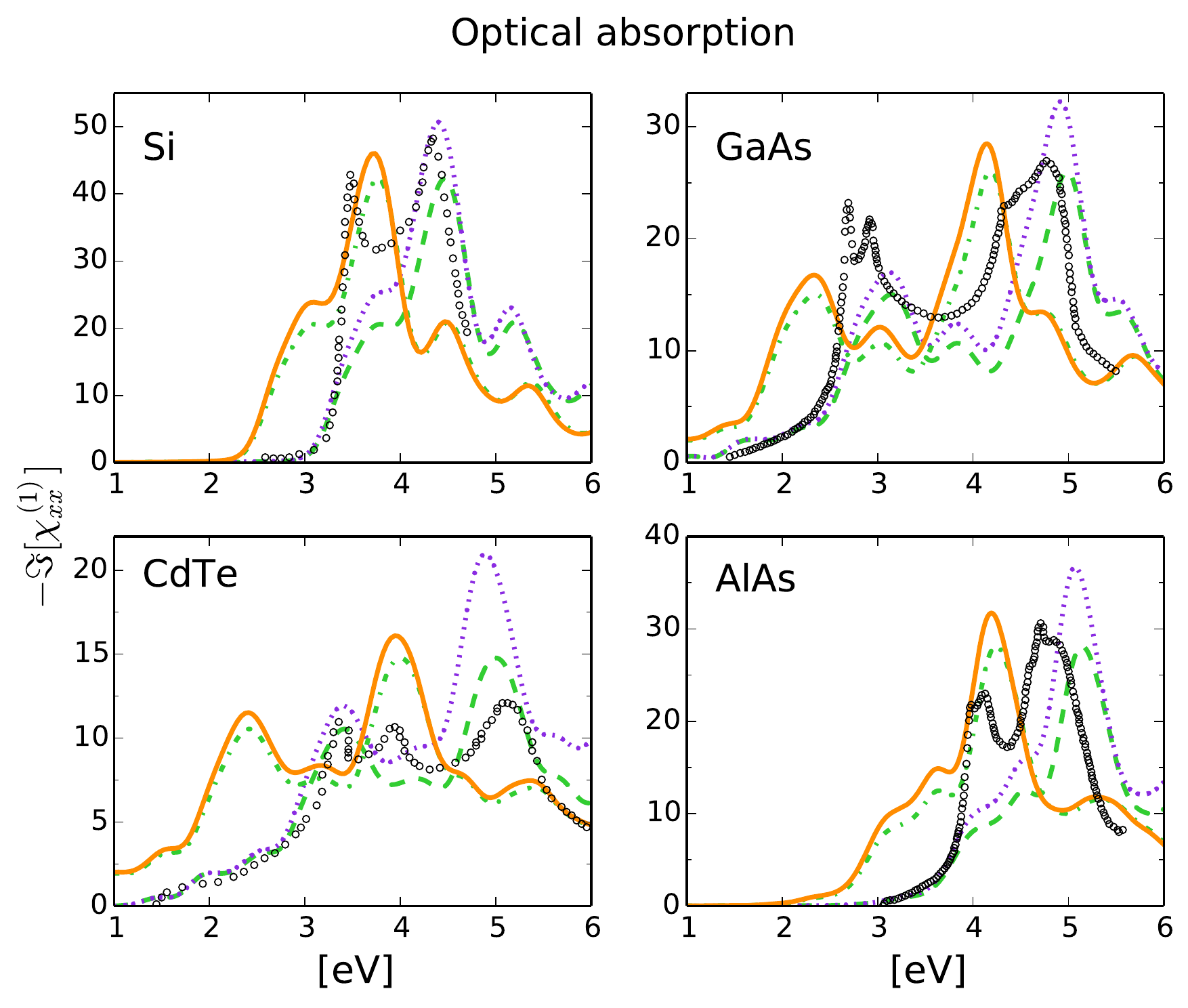}
\caption{\footnotesize{Optical absorption in bulk Si (top left), GaAs (top right), CdTe (bottom left) and AlAs (bottom right): experimental optical absorption spectra (open circles) are compared with real-time simulations at different levels of approximation: TD-LDA (continuous orange line), RPA (green dash-dotted line), both without the scissor correction, and the IPA (violet dotted line) and RPA (green dashed line) with scissor correction.}} \label{fg:epsblk}
\end{figure}

The experimental optical spectra on Si~\cite{PhysRevB.36.4821}, GaAs~\cite{PhysRevB.35.9174}, CdTe~\cite{Adachi} and AlAs~\cite{GARRIGA} (Fig.~\ref{fg:epsblk}, black dashed lines) show qualitative similarities. They all present two main features, a peak at about 3-3.5 eV (referred as $E_1$) and stronger second peak at 4.5-5.0 eV (referred as $E_2$). In GaAs and CdTe, containing heavier third/fourth rows atoms, the $E_1$ peak is split because of the spin-orbit interaction.
Note that we do not include spin-orbit in the KS Hamiltonian and therefore we do not reproduce the splitting at any level of the theory. 

Figure~\ref{fg:epsblk} compares the experimental optical absorption spectra with $-\Im[\chi_{ii}]$ (i.e. the imaginary part of the diagonal of the polarizability tensor, where $i$ is any of the directions $x,y,z$, see App.~\ref{appA})), obtained from the RPA and the TD-LDA (without scissor correction). For the considered systems the two approximations produce very similar spectra. As the only difference between the TD-LDA and the RPA is the microscopic xc potential, one can infer that the effect of the latter is minor as already discussed in the literature.~\cite{botti2007time,Onida2002} 
The most striking difference between the experimental and calculated spectra is the onset that is underestimated by 0.5--1.0~eV. When a scissor operator is added (see Table~\ref{tb:lrcprm}) the agreement is improved though for Si, GaAs and AlAs the $E_2$ peak is slightly blue-shifted and more importantly the $E_1$ peak is either underestimated or appears as a shoulder. Indeed the underestimation of the $E_1$ peak intensity in semiconductor by TD-LDA (and similar TD-DFT approximations) is well known and a signature of missing long-range correlation (see for example Refs.~\onlinecite{PhysRevLett.43.387,PhysRevB.21.4656,botti2007time,Onida2002}). 
Comparison of the RPA spectra and the independent particle approximation (IPA) spectra shows that crystal local fields effects mostly reduces the intensity of the $E_2$ peak by 15--25\%.  
The experimental optical spectrum of CdTe is well caught within the RPA, but for the overestimation of the $E_2$ peak intensity.  

\begin{figure}[t]
\centering
\includegraphics[width=0.5\textwidth]{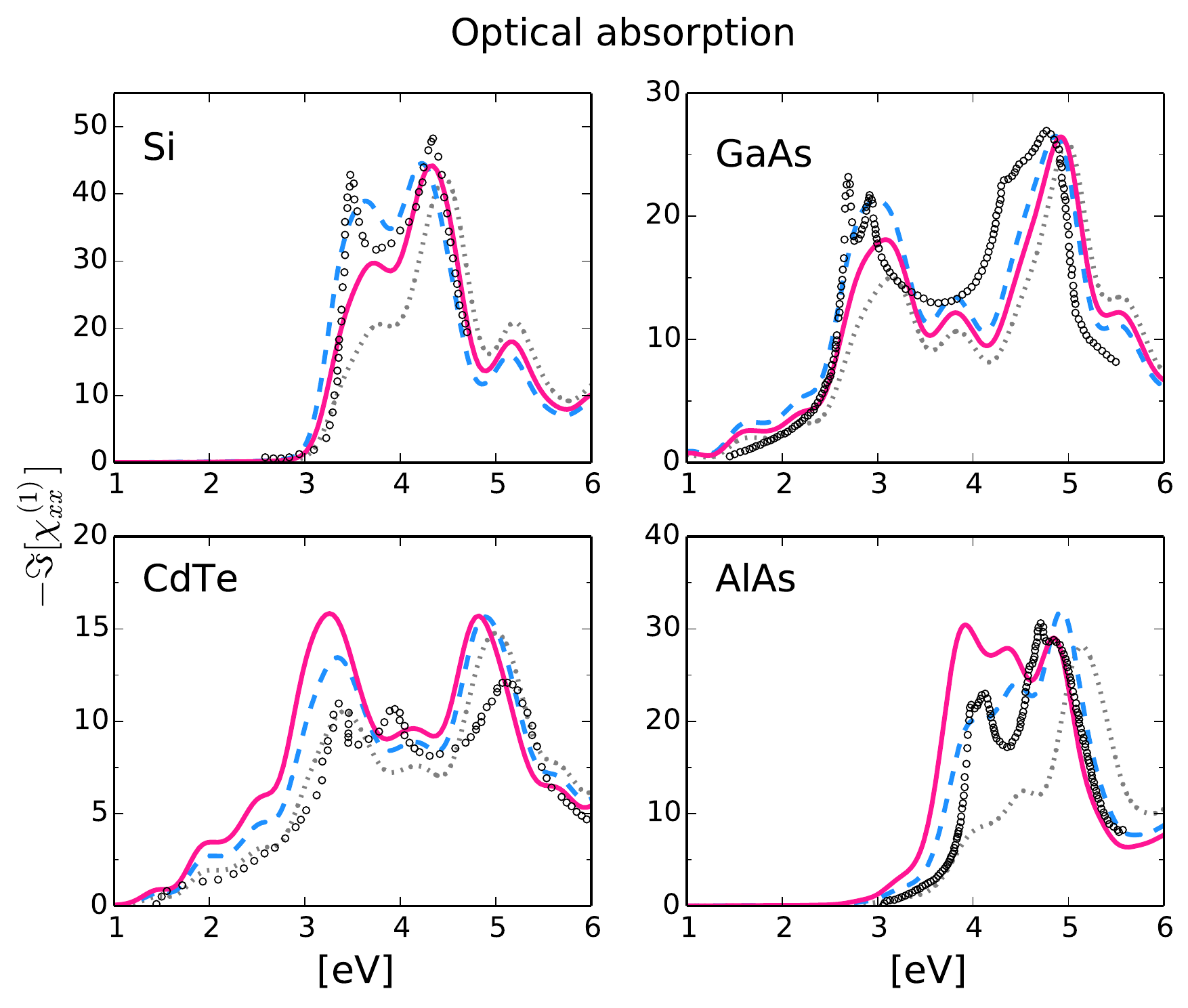}
\caption{\footnotesize{Optical absorption in bulk Si (top left), GaAs (top right), CdTe (bottom left) and AlAs (bottom right): experimental optical absorption spectra (open circles) are compared with real-time simulations at different levels of approximation: opt-PF (blue dashed line), JGM-PF (pink continuous line), RPA(gray dotted line). All approximations include the scissor operator correction.}} \label{fg:epsblk1}
\end{figure}

Figure~\ref{fg:epsblk1} shows the effects of the macroscopic xc field that is added through the approximated PFs discussed in Sec.~\ref{sc:ksef}. For Si, GaAs and AlAs a clear improvement is observed for the opt-PF: both intensity and position of the peaks are reproduced reasonably well. For CdTe adding the xc macroscopic field lead to an overestimation of the $E_1$ peak intensity which was well caught within the RPA. On the other hand the $E_1/E_2$ intensity ratio is better reproduce by the PFs than within RPA.
For the JGM-PF the agreement is in general less satisfactorily. In particular for Si the $E_1$ peak intensity is still visibly underestimated, while for AlAs it is overestimated. The main difference between the two approximation is the value of $\alpha$: in the opt-PF, $\alpha$ is a parameter optimized to reproduce the optical spectra; in the JGM-PF $\alpha$ is determined from the jellium with a gap model. The model does not reproduce the optimal value. For Si, $\alpha^{\sss JGM} \approx 0.11$ and for AlAs $\alpha^{\sss JGM} \approx 0.52$ respectively smaller and larger than the optimal value reported in Table~\ref{tb:lrcprm}.

\begin{figure}[t]
\centering
\includegraphics[width=0.5\textwidth]{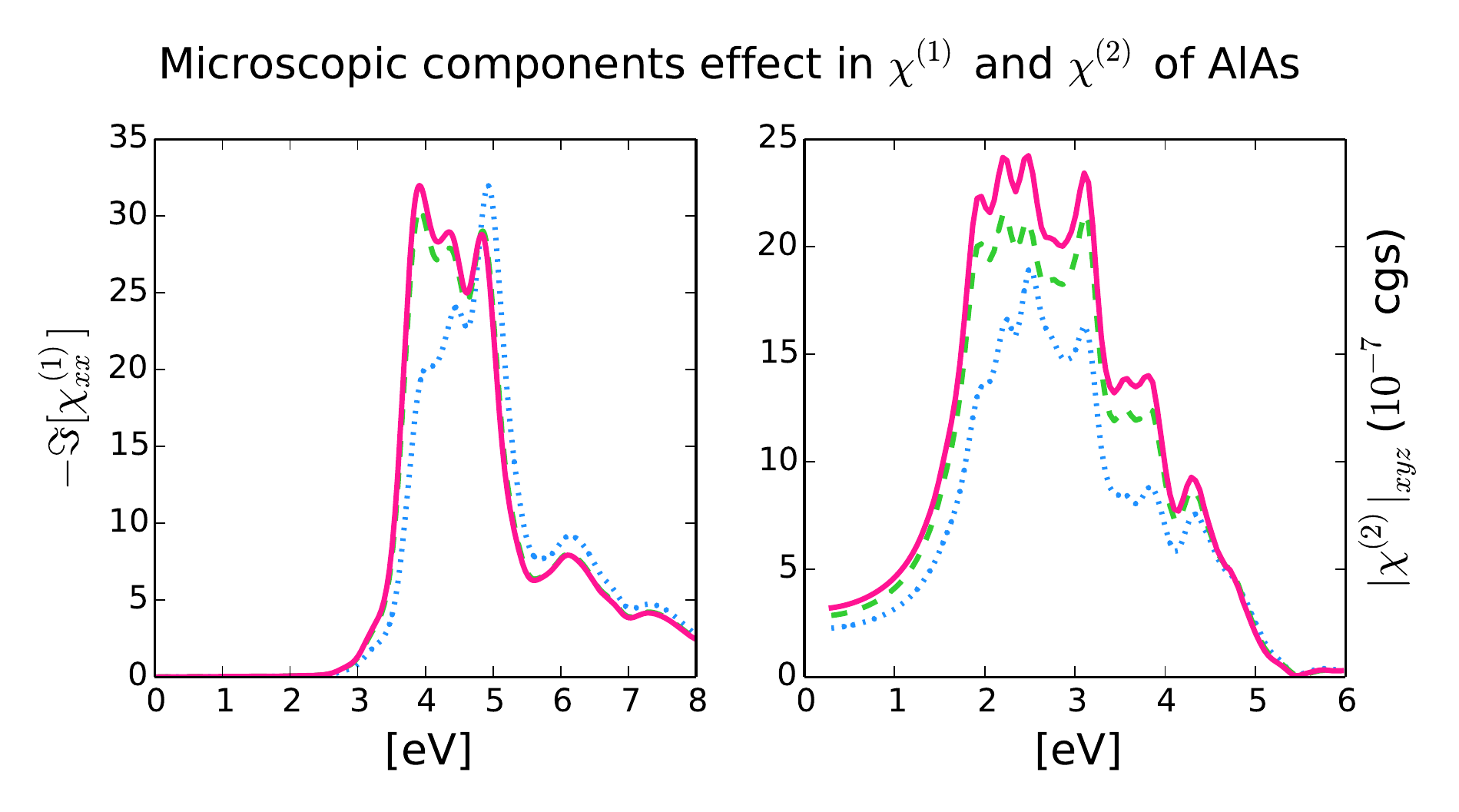}
\caption{\footnotesize{Effect of microscopic components in the JGM-PF on the optical absorption (right panel) and SHG (left panel) of AlAs. The plots compare JGM-PF spectra with (green dashed line) and without microscopic effects (magenta continuous line) and the opt-PF (blue dotted line).}}
\label{fg:effG}
\end{figure}

It is worth to notice that the xc macroscopic field in the JGM-PF has as well a microscopic contribution. For AlAs this contribution is singled out in the right panel of Fig.~\ref{fg:effG} where it is shown to reduce slightly the absorption. For silicon (not shown) the microscopic contribution to the macroscopic field is negligible. Our results for Si and GaAs are slightly different from the one obtained in Ref.~\onlinecite{jgm} though in principle  the magnitude of the applied electric fields is within the linear response limit. In fact differences are expected because of small differences in the numerical parameters of the calculations (e.g the pseudopotential, the $k$-point integration, the broadening). We have verified that when using exactly the same numerical parameters the linear response and the real-time approaches give indeed the same optical absorption spectra for the systems here studied.    

\subsection{Effect of xc macroscopic field on optical absorption} 

\begin{figure}[t]
\centering
\includegraphics[width=0.5\textwidth]{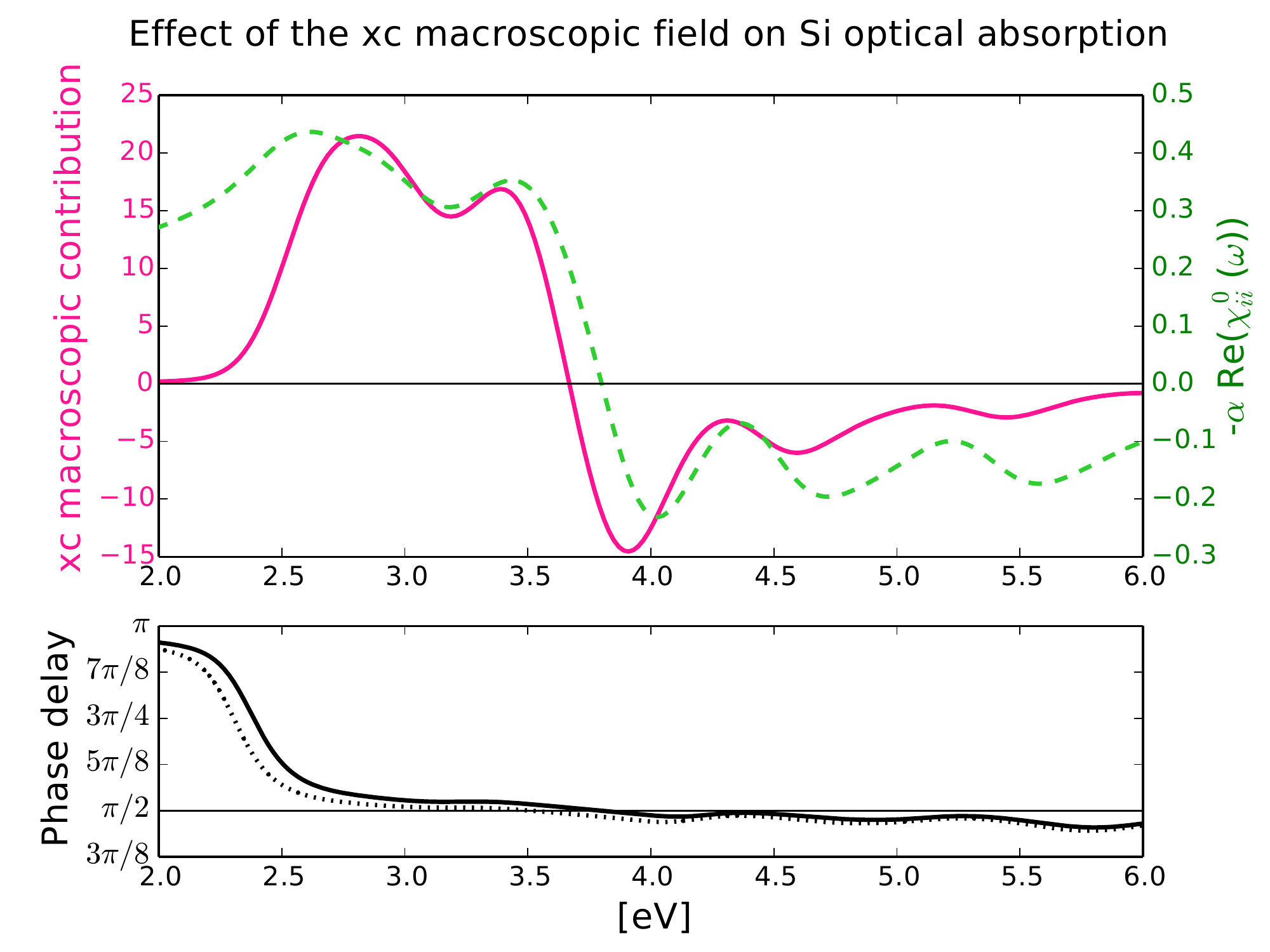}
\caption{\footnotesize{Upper panel: Contribution of the macroscopic xc field to the optical spectrum of Si calculated as the difference between the the opt-PF and the RPA optical absorption spectra (pink continuous line) compared with $-\alpha \Re(\chi_{ii}^0)$ (green dashed line). Bottom panel: phase delay $\phi$ between the polarization and the applied electric field as a function of the applied field frequency at the RPA (dotted line) and opt-PF (continuous line) level of approximation. The horizontal line highlight the $\phi =\pi/2$ delay. See text.}} \label{fg:epsanl}
\end{figure}

It is interesting to analyze how an apparently simple approximation for the xc macroscopic field such as the $\alpha \PP$ (in the opt-PF and JGM-PF) correctly ``distinguishes'' where to increase the optical absorption spectrum at RPA level. 
This information is ``encoded'' in the macroscopic polarization. In fact, in the linear response limit the effective Kohn-Sham electric field within the proposed PF approximations takes the form
$$\Efield^{s} (\omega)= (1 -\alpha\,\newtensor{\chi}(\omega)) \Efield^{\text{tot}} (\omega).$$
That is, the intensity of the applied field is either amplified or reduced depending on the sign of $\Re[\chi_{ii}(\omega)]$
since $\Im[\chi_{ii}(\omega)] \ge 0$ for any positive $\omega$. In Fig.~\ref{fg:epsanl} (upper panel) we see that indeed the sign of $-\Re[\chi_{ii}^0(\omega)]$ (the real part of the RPA macroscopic response function) follows closely that of the correction induced by xc macroscopic contribution $-\alpha \PP$ which has been calculated by subtracting the optical absorption obtained by the RPA,  $\Im[\chi_{ii}^0] $, from the optical absorption obtained by opt-PF,  $\Im[\chi_{ii}] $. 
To gain an insight on how the sign of $\Re[\chi_{ii}^0]$ is linked to the localization of the excitation we consider the phasor representation of $\chi_{ii}^0(\omega) = |\chi_{ii}^0(\omega)|\exp{(i\phi)}$: the complex argument $\phi$ (see bottom panel of Fig.~\ref{fg:epsanl}) gives the phase delay between $\PP$ and $\Efield$. In particular a delay of $\phi = \pi/2$ corresponds to in-phase oscillation of the macroscopic polarization current $\JJ=-\partial \PP/\partial t$ with $\Efield$. Where the optical absorption is negligible those oscillations are plasmons; where instead it is non--negligible they can be considered as a signature of delocalized excitations (note that in fact the optical absorption has a maximum at $\phi=\pi/2$).  Heuristically, for more localized excitations we may expect a phase delay larger than $\pi/2$, and for delocalized excitations a phase delay smaller than $\pi/2$.

Then, the $\cos\phi$, and $\Re(\chi_{ii})$ which is proportional to it, are negative for localized excitations and positive for the more delocalized ones. A correction proportional to  $-\Re(\chi_{ii})$ then increases the absorption in correspondence of more localized excitation and decreases it for more delocalized excitations. Note as well that in the RPA the phase delay is overestimated. Then the absorption, proportional to $\sin\phi$ is too small for $\phi > \pi/2$ (localized excitation) and too large for $\phi < \pi/2$ (delocalized  excitation).

\subsection{Second-harmonic generation of GaAs, AlAs and CdTe} 

\begin{figure}[t]
\centering  \includegraphics[width=0.5\textwidth]{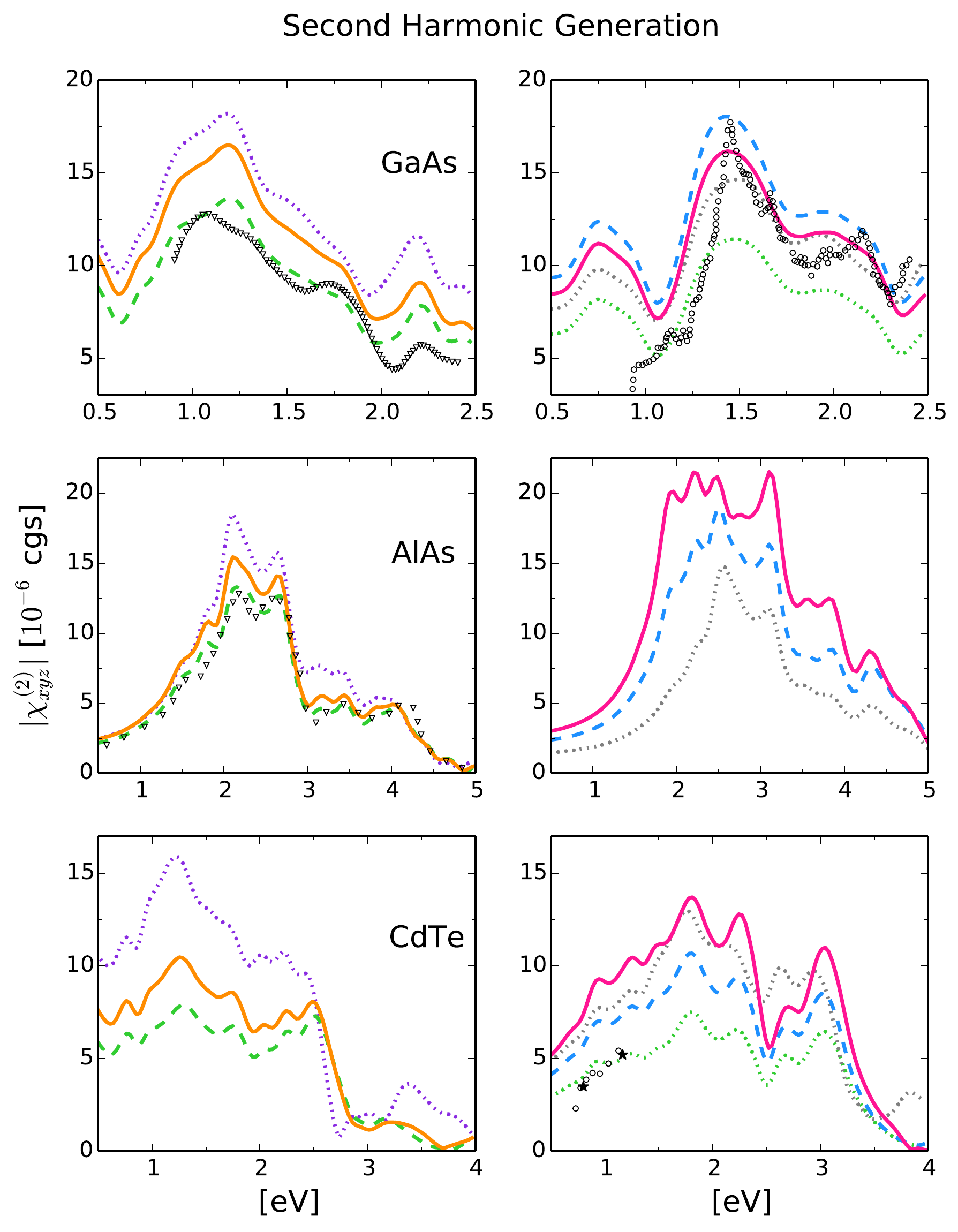}
\caption{\footnotesize{SHG spectra of GaAs (top panels),  AlAs (middle panels) and CdTe (bottom panels) obtained from real-time simulations at different levels of approximation. Left panels: IPA (dotted violet), RPA (dashed green) and TD-LDA (continuous orange)---all without scissor operator correction. For comparison we included the RPA spectrum of GaAs and AlAs calculated by Luppi et al.\cite{PhysRevB.82.235201}} (open triangles). Right panels: opt-PF (dashed blue) and JGM-PF (continuous pink) are compared with IPA (dotted gray) and RPA for CdTe and GaAs. Available experimental results are shown for GaAs (open circles)~\cite{bergfeld2003second} and CdTe (open circles~\cite{Shoji:97} and stars~\cite{Jang:13}).\label{fg:shg}}
\end{figure}

In zincblende structures the only independent non-zero SHG component~\cite{Boyd20081} is $\chi^{(2)}_{xyz}$ (or its equivalent by permutation). The module of the calculated $\chi^{(2)}_{xyz}$ for the systems under study is reported in Fig.~\ref{fg:shg} and compared with experimental values where available.
Note that when the energies are not corrected by a scissor (left panel) for both GaAs and CdTe a large part of the energy range of the SHG spectra is in the absorption region where both one-photon and two-photon resonances contribute to the intensity.  For AlAs the part of the SHG spectra below 2~eV is instead in the transparency region of the material (only two-photon contributions).
When the scissor-correction to the energy is applied (right panel), the transparency region for GaAs and CdTe is below 1~eV and for CdTe below 3~eV. In the transparency region only two-photon resonances contribute. 
Comparing the TD-LDA with the RPA and the independent particle (IP)  spectra (left panel) shows that crystal local field effects (that tend to reduce the overall SH intensity) are partially compensated by the microscopic xc effects (that tend to increase the SH intensity). In general both effects are relatively stronger than for the optical absorption. Applying the scissor correction does not correspond to a simple shift (like in the optical absorption case) but changes the spectra. Firstly the SH intensity is reduced overall (because of sum rules), secondly the intensity is redistributed as the scissor modifies the relative position of one-photon and two-photon resonances (that are shifted by a half of the scissor value). For GaAs and CdTe the addition of macroscopic correlation through the approximated PF leads to an enhancement of about 40\% in GaAs and 80\% in CdTe with respect to the RPA. On the other hand as discussed for those systems local field effects are very large and in fact the spectra form the PF are not significantly different than at the IP level, meaning an almost exact cancellation of the crystal local effects and the macroscopic xc effects as describe by the approximated PFs. Only in the case of AlAs, the macroscopic correlation enhances significantly the SH, adds features and redistributes relative weights with respect to the IP approximation. 

Regarding the comparison with experiment (right panel), in GaAs the peak at 1.5~eV and the feature at 2.2~eV in the experimental SHG are fairly reproduced by the opt-PF and  JGM-PF approximations. All approximations significantly overestimate the SHG for energies below 1~eV. A similar breakdown of the opt-PF approximation (that within the response theory context corresponds with the long-range corrected kernel) has been observed by Luppi and coworkers and traced back to the errors in the theoretical macroscopic dielectric function.~\cite{Luppi2010} For CdTe, the approximation that is closer to experimental results (which however are available only around 1~eV) is the RPA while both PF approximations overestimate the experimental SH. This is consistent with the results for optical absorption for which the RPA gives the best agreement among all approximations considered.

We have also compared our results from real-time simulations with those obtained from a response approach by Luppi and co-workers~\cite{Luppi2010} and we found a good agreement, slightly better than our previous work\cite{nloptics2013} thanks to the higher order approximation for the covariant derivative [Eq.~\eqref{eq:wkhat2}]. In the left panel of Fig.~\ref{fg:shg} we show for example the comparison for the RPA. There is a very good correspondence between the two spectra for AlAs. For GaAs there are small, but still visible differences which we argue are due to the different pseudopotentials used. In fact we obtain a similar variation in our results when repeating the calculations with different pseudopotentials. It is known that SHG is very sensitive to changes in the electronic structure and that is turn changes when using different pseudopotentials. This is particularly true in the case of GaAs and the sensitivity on the pseudopotential choice was also observed in the referenced calculations.  
Note that in the pseudopotentials we used $d$ orbitals are considered as core electrons, whereas they are included as valence electrons in the calculation of Luppi and coworkers.~\cite{Luppi2010} On the other hand pseudopotentials including $d$ electrons that we were testing did not provide a much better agreement.  

\subsection{Third-harmonic generation of Si} 

\begin{figure}[t]
\centering
\includegraphics[width=0.5\textwidth]{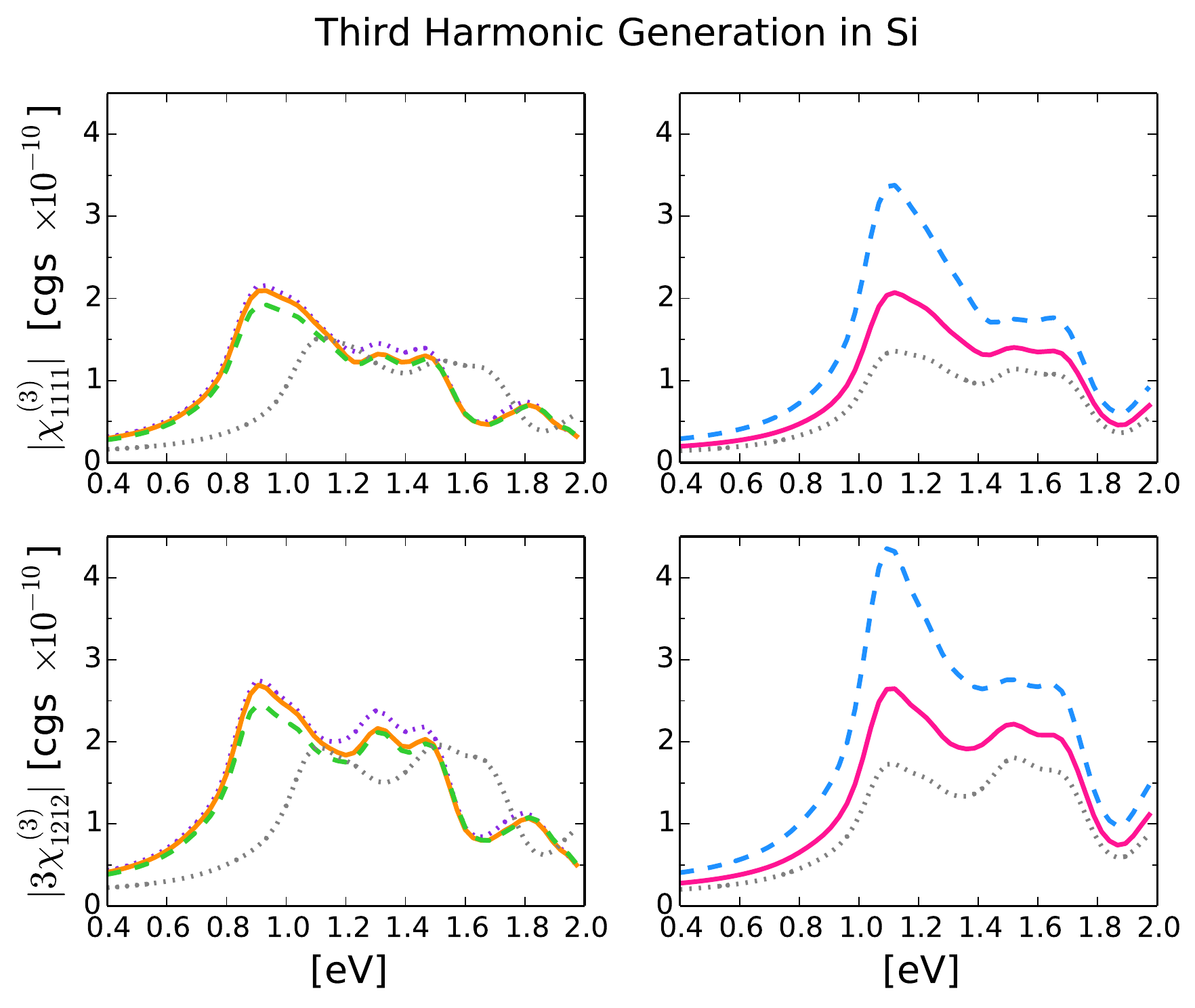}
\caption{\footnotesize{THG of Si: $|\chi^{(3)}_{1111}|$ and $|3\chi^{(3)}_{1212}|$ components (see text). Spectra obtained from real-time simulations at different levels of approximation. Left panels: TD-LDA (continuous orange line), RPA (green dashed line), IPA (dotted violet line) without scissor operator correction are compared with and IPA (gray dotted line) with scissor operator correction. Right panels: JGM-PF (continuous pink line), opt-PF (blue dashed line) and RPA (gray dotted line) with scissor operator correction.}} \label{fg:siX3ab}
\end{figure}

 The THG for Si has two independent components, $\chi^{(3)}_{1212}  \equiv \chi^{(3)}_{xyxy}$ and $\chi^{(3)}_{1111} \equiv \chi^{(3)}_{xxxx}$.  In the expression for the TH polarization along the direction $i$,
$$P_i(3\omega) = 3\chi^{(3)}_{1212} \efield_i(\omega)|\efield(\omega)|^2+(\chi^{(3)}_{1111} - 3\chi^{(3)}_{1212}) \efield_i^3(\omega),$$
$3\chi^{(3)}_{1212}$ is the isotropic contribution, while $\chi^{(3)}_{1111}$ the anisotropic contribution. Figure~\ref{fg:siX3ab} shows the calculations for $A=|\chi^{(3)}_{1111}|$ and $B=|3\chi^{(3)}_{1212}|$, the modules of the $1111$ and $1212$ components of the THG of Si.~\cite{Moss:89} 
The TD-LDA spectra (left panels) both present two main features, a peak around 0.9~eV (three-photon resonance with $E_1$) and a shoulder around 1.4  eV (three-photon resonance with $E_2$). Both features are more intense and pronounced in the $|3\chi^{(3)}_{1212}|$. Results within TD-LDA resemble closely those obtained within the RPA and IP approximation. For the $E_1$ three-photon resonance the microscopic xc effects cancel with the local-field effects, so that TD-LDA almost coincides with the IP approximation. For higher energies instead, the TD-LDA and RPA spectra are practically identical. Applying a scissor operator does not simply shift the peaks by an amount of about 1/3 of the scissor value. The overall intensity of the spectra is reduced (as expect from sum rules) and as well the relative intensity of the $E_1$/$E_2$ resonances changes. Specifically the ratio is close to or even smaller than $1$ in the scissor corrected spectra, while is $\approx 1.2-1.3$ in the uncorrected spectra. The macroscopic xc field introduced with the approximations for the PF (right panels) enhances the intensity of the spectra and as well the $E_1$/$E_2$ ratio. Consistently with what observed for the linear response the largest $\alpha$ (opt-PF for silicon) produces the largest correction.

\begin{figure}[t]
\centering
\includegraphics[width=0.5\textwidth]{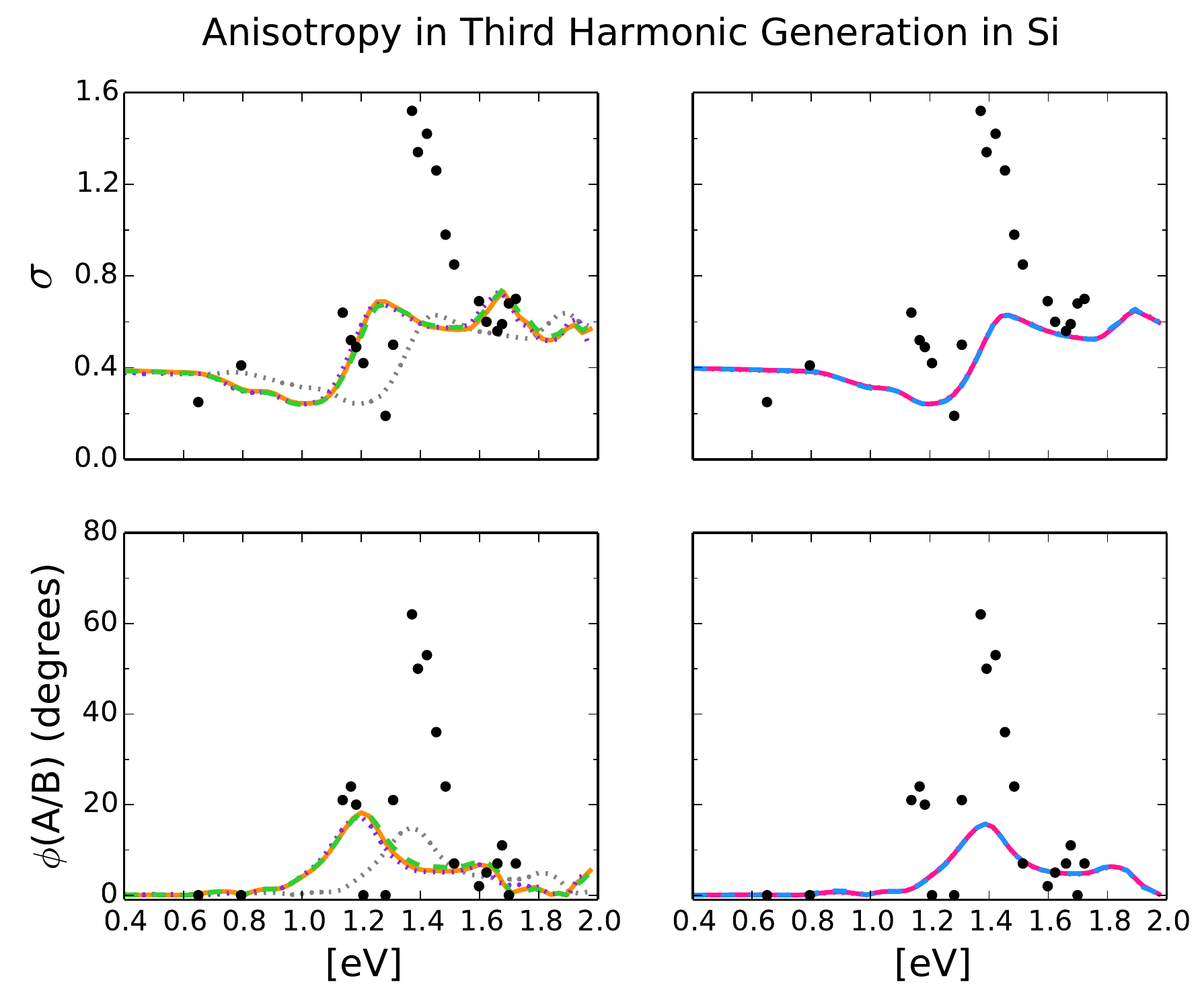}
\caption{\footnotesize{THG of Si: anisotropy parameters $\sigma$ and $\phi$ (see text). Experimental data (open circles)\cite{Moss:89} compared with results obtained from real-time simulations at different levels of approximation. Left panels: TD-LDA (continuous orange line), RPA (green dashed line), IPA (dotted violet line) without scissor operator correction are compared with the IPA (gray dotted line) with scissor operator correction. Right panels: JGM-PF (continuous pink line), opt-PF (blue dashed line) and RPA (gray dotted line) with scissor operator correction.}} \label{fg:siX3an}
\end{figure}
Experimental measurements are available for the ratio $R_1$ between the THG signal obtained with $45$ and $0$ incident angles and for the ratio $R_2$ between the THG signal obtained with circularly polarized light and linearly polarized light at $0$ incident angle. From those measurements then $\sigma = |1 - B/A|$ and the phase $\phi(B/A)$ can be deduced.~\cite{Moss:89} The experimental results are reported in Fig.~\ref{fg:siX3an}. Both $\sigma$ and $\phi(A/B)$ present two features at about 1.1~eV and 1.4~eV in correspondence of the three-photon $E_1$ and $E_2$ resonances. All the theoretical results are very similar irrespective of the approximation used and the differences observed for the $A=|\chi^{(3)}_{1111}|$ and $B=|3\chi^{(3)}_{1212}|$ in Fig.~\ref{fg:siX3ab}. The results from the scissor corrected approximations (right panels) are just shifted by 1/3 of the scissor operator. When compared with the experiment all the approximation reasonably reproduce the behavior at energies lower than 1~eV. However for both $\sigma$ and $\phi(A/B)$ (we consider here only the scissor corrected approximations which have resonances at the correct energies) the peak in correspondence of the $E_1$ resonance is missing and the feature in correspondence of the $E_2$ resonance much less pronounced than in experiment. When compared with calculations from Moss and coworkers~\cite{Moss1990} at the independent particle level from the electronic structure calculated either with empirical tight-binding and semi-ab-initio band-structure techniques, the intensity we found for $A$ and $B$ are similar to the latter, but the main spectral features are similar to the former. To notice that the THG based on empirical tight-binding shows in the $\sigma$ and $\phi$ spectra a peak at 1.1~eV.

\section{Summary and conclusions}
We have implemented a real-time density functional approach suitable for infinite periodic crystals in which we work within the so-called length gauge and calculate the polarization as a dynamical Berry phase.~\cite{souza_prb}        
This approach, in addition to the electron density considers also the macroscopic polarization as a main variable and extends to the time-dependent case the DPFT introduced in the nineties~\cite{Gonze1995,Resta1996,Vanderbilt1997,Martin1997} to correctly treat IPC in electric fields within a density functional framework. In the corresponding time-dependent KS equations next to the microscopic xc potential also appears a macroscopic xc electric field which is a functional of the macroscopic polarization (and eventually of the microscopic density).
We have derived approximations for the xc electric field exploiting the connection with long-range corrected approximations for xc kernel within the linear response theory. We have considered two approximations, the optimal polarization functional, linked to the  long-range corrected xc kernel proposed on Ref.~\onlinecite{botti2004long} and the Jellium with a gap model polarization functional linked to the analogous approximation for the xc kernel.\cite{jgm}
We have applied this approach, that we refer to as real-time DPFT, to calculate the optical absorption, second and third harmonic generation in different semiconductors (Si, GaAs, AlAs and CdTe). We have compared results with ``standard'' real-time TD-DFT, namely without macroscopic xc effects, and to experimental results where available. The general trend is an overall improvement over standard TD-DFT as to be expected from the results obtained within the response framework.~\cite{botti2004long} Of the considered approximations, the opt-PF provides the best agreement with the experiment.

The approach here proposed combines the flexibility of a real-time approach, with the efficiency of DPFT in capturing long-range correlation. It allows calculations beyond the linear regime (e.g. second- and third-harmonic generation, four-wave mixing, Fourier spectroscopy or pump-probe experiments) that includes excitonic effects. It is an alternative approach to real-time TD-DFT for extended system proposed by Bertsch, Rubio and Yabana.\cite{PhysRevB.62.7998} At variance with our approach the latter uses the velocity gauge---which has the advantage of using the velocity operator that is well defined in periodic systems---rather than the position operator that requires special attention. On the other hand,  although this approach have shown promising results,\cite{yabana2012time,goncharov2013nonlinear} it turns to be quite cumbersome for studying response functions beyond the linear regime due to the presence of divergences that in principle should cancel, but that are difficult to treat numerically.\cite{PhysRevB.52.14636} Furthermore non-local operators---such as pseudo-potentials or the scissor operator---are cumbersome to treat in velocity gauge\cite{tokman} while they are trivial in length gauge.

Similarly to any density-functional approaches, a delicate point is the approximation of the xc effects. In addition to the xc potential as in standard DFT, in this approach we also need an approximation for the macroscopic xc field. Though for the systems here studied the opt-PF approximation seems to work well, such a good performance cannot be expected in general. For example, based on the experience from linear response calculations, this approximation is not expect to work very well for large gap insulators or systems with a reduced dimensionality (e.g. nanostructures or layers) in which the electronic screening is small.~\cite{PhysRevB.68.205112} Furthermore, in the opt-PF the $\alpha$ is chosen has a material dependent parameter rather than obtained from first-principles. In this respect within the other approximation here studied, JGM-PF, $\alpha$ is determined from first-principles but not always has the optimal value. Further studies then should try to develop universal approximations to the polarization functional, possibly going beyond the linear response formulation that was here used in the derivation of the polarization functionals.

\section{Acknowledgments}
\label{ackno}                                        
This work used the computing facilities of the Atomistic Simulation Centre--Queen's University Belfast, of the CINaM Aix-Marseille Universit\'e, of the ARCHER UK National Supercomputing Service (http://www.archer.ac.uk)  through EPSRC grant EP/K0139459/1 allocated to the UKCP Consortium, and of the ``Curie" national GENGI-IDRIS supercomputing center under contract No. x2012096655.
CA acknowledges EUSpec Cost Action MP1306. DS acknowledges the {\em Futuro in Ricerca} grant No. RBFR12SW0J of the
Italian Ministry of Education, University and Research MIUR, 
the European Union project
MaX {\em Materials design at the eXascale} H2020-EINFRA-2015-1, Grant agreement n. 676598 and
{\em Nanoscience Foundries and Fine Analysis - Europe} H2020-INFRAIA-2014-2015,
Grant agreement n. 654360.

\appendix
\section{Induced field and response functions} \label{appA}

One of the objectives of atomistic simulations is the calculation of
the macroscopic dielectric function or of related response functions of dielectrics.
Within TD-DFT such goal is achieved via the calculations of the
microscopic density--density response function $\tchirr$, defined via the equation
\be
\delta n_{\sss \GG}(\qq,\w)=\tchirr_{\sss \GG\GG'} (\qq,\w)\ \delta v^{\text{ext}}_{\sss \GG'} (\qq,\w).
\label{eq:chi_red} 
\ee
Here $\GG$ are the reciprocal lattice vectors and $\w$
the frequency obtained from the Fourier transforms $\rr\ra\GG$ and $t\ra\w$.
In addition to  $\tchirr$, 
 the irreducible response function $\chirr$ and
the auxiliary response function $\bchirr$ can be defined via
\bea
\delta n_{\sss \GG}(\qq,\w)&=&\chirr_{\sss \GG\GG'}(\qq,\w)\
 \delta v^{\text{tot}}_{\sss \GG'}(\qq,\w) \label{eq:chi_irr} \\
\delta n_{\sss \GG}(\qq,\w)&=&\bchirr_{\sss \GG\GG'} (\qq,\w)
 [\delta v^{\text{ext}}_{\sss \GG'}(\qq,\w)+\delta \bar v^\text{H}_{\sss \GG'}(\qq,\w)]. \label{eq:chi_bar}
\eea

To linear order and at finite momentum (i.e. $\qq\neq\zero$),
the longitudinal microscopic dielectric function can be derived from the response functions,
\bea
\epsilon^{-1}_{\sss \GG\GG'}(\qq,\w)=\delta_{\sss \GG,\GG'}
  + 4\pi \frac{\tchirr_{\sss \GG\GG'}(\qq,\w)}{|\qq+\GG||\qq+\GG'|}
\label{eq:eps_M1_micro} , \\
\epsilon_{\sss \GG\GG'}(\qq,\w)=\delta_{\sss \GG,\GG'}
  - 4\pi \frac{\chirr_{\sss \GG\GG'}(\qq,\w)}{|\qq+\GG||\qq+\GG'|}
\label{eq:eps_micro}
\text{.}
\eea
The longitudinal macroscopic dielectric function can then be obtained as
$\epsilon_M(\qq,\w)=1/\epsilon^{-1}_{\sss \zero\zero}(\qq,\w)$.
Absorption experiment however are described at $\qq=\zero$
where the dielectric function 
$\epsilon_M(\w)\equiv \epsilon_M(\zero,\w)$
can be obtained only via a limiting process.
They are defined as
\bea
\epsilon_M(\w)&=&\left[ 1 + 4\pi\lim_{\qq\ra 0} \frac{\tchirr_{\sss \zero\zero}(\qq,\w)}{|\qq|^2}\right]^{-1} \\
\epsilon_M(\w)&=&\      1 - 4\pi\lim_{\qq\ra 0} \frac{\bchirr_{\sss \zero\zero}(\qq,\w)}{|\qq|^2}
\text{.}
\eea
As we observed in the introduction this approach is at least problematic in real-time simulation,
where it is numerically more convenient to directly work at $\qq=0$ and thus the density--density
response function cannot be used.

Within DPFT the key quantity is the one which relates the macroscopic electric
field $\Efield^{tot}$ or $\Efield^{ext}$ to the first order polarization $\PPo 1$.
\bea
\PPo 1(\w)=\newtensor{\tilde{\chi}}(\w) \Efield^{ext}(\w) \label{eq:tsusc1} \\
\PPo 1(\w)=\newtensor{       \chi }(\w) \Efield^{tot}(\w)  \label{eq:susc1}     
\text{.}
\eea
$\newtensor{\chi}(\w)=\newtensor{\chi}^{(1)}(\w)$ is the (first--order) polarizability;
$\newtensor{\tilde{\chi}}(\w)=\newtensor{\tilde{\chi}}^{(1)}(\w)$ is the quasi--polarizability.
Since we obtain the polarizability dividing the Fourier transform of the time--dependent
polarization by the input electric field, we obtain either
$\newtensor{\tilde{\chi}}(\w)$ or $\newtensor{\chi}(\w)$
depending on whether we assume $\Efield^{\text{inp}}=\Efield^{\text{ext}}$ or $\Efield^{\text{inp}}=\Efield^{\text{tot}}$.
Notice that in this framework we have already made the distinction between macroscopic fields,
described in terms of $\Efield^{\text{ext}}/\Efield^{\text{tot}}$, and microscopic ones,
described in terms of $\bar v^{\text{tot}}/\bar v^{\text{tot}}$.
$\newtensor{\tilde{\chi}}(\w)$ and $\newtensor{\chi}(\w)$
are thus macroscopic functions.
The longitudinal dielectric function can be obtained, to first order in the field, as
\bea
\epsilon_M(\w) &=& \left[ 1 + 4\pi\tilde{\chi}_{ii}(\w)\right]^{-1}, \\
\epsilon_M(\w) &=&\       1 - 4\pi       \chi _{ii}(\w),
\eea
where $\tilde{\chi}_{ii}$ is any of the diagonal components of $\newtensor{\chi}$.

More in general the $n$-order polarization can be expressed as
\begin{multline}\label{eq:PpwrfE}
\PPo n(t)= \int dt_1\ ...\ dt_n\times \\ \susc n(t-t_1,\ ...\ ,t-t_n)\times  \\ 
         \Efield^{\text{tot}}(t_1)\ ...\ \Efield^{\text{tot}}(t_n) ,
\end{multline}
where $\susc n$ is the $n$-order polarizability related to $n$-order nonlinear
optical properties.
Also here we could define the $\tsusc n$ as the response to the external field.
The two can be related from the equation
\be
\tsusc n(\w)=\susc n(\w)(1-4\pi\susc 1)^n
\ee
As for the linear case we obtain either $\tsusc n(\w)$ or $\susc n(\w)$
depending on whether we assume $\Efield^{inp}=\Efield^{ext}$ or $\Efield^{inp}=\Efield^{tot}$.
However, since usually $\susc n(\w)$ is the quantity considered in the literature the
last choice is more convenient in nonlinear optics.

\addcontentsline{toc}{chapter}{Bibliography}
\bibliographystyle{apsrev4-1}

\begin{thebibliography}{70}
\bibitem [{not({\natexlab{c}})}]{notepath}%
  \BibitemOpen
  \href@noop {} {} \bibinfo {note} {In general the
  expression of the xc electric field in terms of $\newtensor{F}^{\sss XC}_{}$
  is an integral along a path in the infinite dimensional space of the
  densities/macroscopic polarizations. Such integral does not depend on the
  path if $\newtensor{F}^{\sss XC}_{}$ is defined as a functional derivative of
  some function. Furthermore here we consider the case where
  $\newtensor{F}^{\sss XC}_{}$ is a local (or semi--local) functional. Then the
  infinite dimensional integral reduces to a simple three-dimensional integral
  which, in reciprocal space, can be represented as a sum over the $\GG$
  vectors.}\BibitemShut {Stop}%
\bibitem [{not({\natexlab{d}})}]{noterev}%
  \BibitemOpen
  \href@noop {} {} \bibinfo {note} {Note that in
  Refs.~\onlinecite{PhysRevLett.102.113001,Nazarov2010} the scalar kernel is
  instead derived starting from the tensor kernel.}\BibitemShut {Stop}%
\end{thebibliography}
%

\end{document}